\documentclass[twocolumn, secnumarabic, amssymb, nobibnotes, superscriptaddress, aps, prl]{revtex4-1}
\usepackage{amsmath,graphicx,float}

\begin{document}
\bibliographystyle{unsrt}

\title{Description of Brownian motion including both kinetic and hydrodynamic effects}


\author{Hanqing Zhao}
\affiliation{Department of Physics and Institute of Theoretical Physics 
and Astrophysics, Xiamen University, Xiamen 361005, Fujian, China}
\affiliation{Department of Modern Physics, University of Science and Technology of China, Hefei 230026, China}
\author{Hong Zhao}
\email{zhaoh@xmu.edu.cn}
\affiliation{Department of Physics and Institute of Theoretical Physics
and Astrophysics, Xiamen University, Xiamen 361005, Fujian, China}


\begin{abstract}
It is known that a full description of Brownian motion in the entire course of time
should incorporate both kinetic and hydrodynamic effects, but a formula accounts for both effects has
been established only in three dimension and only for the limiting case when Brownian particles are
much bigger and heavier than fluid particles. Nevertheless, for applications, it is important to
consider small Brownian particles (sometimes even smaller than fluid particles). In these cases, unfortunately,
only formulae in the short (long) time limit considering exclusively kinetic (hydrodynamic) effect are available. Here we derive a general solution to
both two- and three-dimensional Brownian motion. Our solution is applicable to a wide spectrum of
suspended particles from being smaller than the fluid particles to usual
Brownian particles. Our analytical results are well corroborated by the numerical simulations.

\end{abstract}
\maketitle

Brownian motion has served as a pilot of studies in diffusion and other transport phenomena for
over a century and has played a vital role in various fields~\cite{resi, peter, 111}. For a suspended particle in a fluid, it has been known that its velocity autocorrelation function (VACF) decays exponentially as
$C_K(t) = C(0) \exp[-C(0)t/D_0]$ at short times~\cite{einstein,uhlen,martin}, followed by a power-law tail $C_H(t)\sim t^{-d/2}$, where $d$ is the fluid dimension~\cite{alder1, alder2, cohen1, cohen2, ernst1, ernst2, zwanzig1, widom, hinch, zwanzig2, russel, new2}. The former is a result of the kinetic theory, while the latter is predicted by hydrodynamics. The hydrodynamic effect may play a non-negligible role even in short times. To get an accurate evaluation of the hydrodynamic effect, by which one can separate the kinetic and hydrodynamic contributions, we need a full expression of the VACF applicable in the entire course of time. Such an expression has been found by the generalized Langevin equation approach based on the fluctuating hydrodynamics~\cite{martin, russel, zwanzig2}. However, it is applicable only for Brownian particles much heavier and bigger than the fluid particles. Significant deviation appears for light Brownian particles as $t\to 0$ due to the incompressibility condition adopted~\cite{martin}. This is a serious limitation since light Brownian particles represent the canonical Brownian motion, as a pollen in water. Moreover, this formula can not be applied to suspended particles with sizes comparable to fluid particles. The diffusion of small particles are of particular interests for applications, such as for small-molecule drugs~\cite{m1,m2}.  Another limitation is that it applies only to three dimension (3D). Nowadays, two-dimensional (2D) fluids have become of practical importance, e.g., 2D water confirmed by graphene
flakes~\cite{g1, g2}, quark-gluon plasma under certain conditions~\cite{q1}, cell membranes and other kind of biomembranes, and so on. Hence a full description for 2D Brownian motion is also desired.

On the other hand, owing to the state-of-the-art technologies~\cite{tech1, tech2, tech3, tech4, tech5, tech6, new1, new3}, in recent years there has been a renewal of experimental interests in examining the particle diffusion theory~\cite{ver1, ver2, ver3, tech4, tech5, ver4}. A focused topic is to probe diffusion anomalies of hydrodynamic origin~\cite{ver1, ver2, tech4, ver4}. As such, a general formula of Brownian motion,  that can clearly distinguish the kinetic and hydrodynamic contributions, becomes urgently important.

The purpose of this work is to derive a general formula for particle diffusion that applies under a variety of conditions exemplified above. We start from the generalized hydrodynamics which suggests to replace the diffusion constant, $D_0$, by a time-dependent coefficient, $D(t)$~\cite{modi,pt}. Consequently, the density distribution becomes $\psi(\mathbf{r},t)=\frac{1}{4\pi D(t)t}\exp[-\frac{r^{2}}{4D(t)t}]$, when the hydrodynamic effect is taken into account. $D(t)$ is connected to the VACF by
\begin{equation}
D(t)=\int_{0}^{t}C(t^{\prime})dt^{\prime},
\end{equation}
which recovers the usual Green-Kubo formula at $t\rightarrow \infty$.

Let us consider a tagged Brownian particle of mass $M$. Without loss of generality, we assume that initially it is located at the origin and moves alone the $x$-axis with a momentum $\mathbf{p}(0)\equiv[p_{x}(0),0,0]$. The VACF is $C(t)=\langle p_x(t)p_x(0)\rangle/M^{2}$, where $\langle\cdot\rangle$ represents the ensemble average.
The instant momentum of the tagged particle can be decomposed into three parts, dubbed the kinetic, hydrodynamic, and random part, respectively: $p_x(t)=p^K_x(t)+p^H_x(t)+p^R_x(t)$. The kinetic part, $p^K_x(t)$, is the portion of $p_{x}(0)$ that remains at time $t$; The hydrodynamic part, $p^H_x(t)$, is the portion that is transferred back to the tagged particle through surrounding particles. The random part, $p^R_x(t)$, comes from random collisions with other particles, which is uncorrelated with $p_{x}(0)$. Taking these into account, we have
\begin{equation}
C(t)=C_{K}(t)+C_{H}(t).
\end{equation}
By the kinetic theory, $p_x^{K}(t)=p_x(0)\exp[-\frac{C(0)}{D_{0}}t]$, which gives $C_K(t)$ straightforwardly, our task is thus to calculate $p^H_x(t)$ to obtain $C_H(t)$. Note that at time $t$, the portion of momentum $p_{x}(0)$ that has been transferred into fluid in the kinetic process is $p_x(0)[1-\exp{(-\frac{C(0)}{D_{0}}t)}]$. Denote by $p_x(\mathbf{r},t)$ its density, i.e., the momentum transferred to a unit volume at position $\mathbf{r}$ and time $t$, then a  fluid particle in the volume has an average velocity $p_x(\mathbf{r},t)/\rho$, which gives the velocity field of vortex backflow triggered by the tagged particle. Here $\rho$ is the fluid density. The tagged particle, irrespective of its shape and mass, has the same velocity as the fluid particle on average if it stays at $\mathbf{r}$ and time $t$  because it is driven by the backflow. This is equivalent to the general assumption that local equilibrium can be established rapidly~\cite{sl}.
Nevertheless, the transferred momentum can not establish the backflow promptly. To determine the time after which the backflow  sets in, we introduce a function $R(t)$ to characterize the retarded effect in response. It leads to
\begin{equation}
p_x^{H}(t)=\frac{M}{\rho}(1-e^{-\frac{C(0)}{D_{0}}t})R(t)\int p_x(\mathbf{r},t)\psi(\mathbf{r},t)d\mathbf{r}.
\end{equation}

\begin{figure}[t]
\includegraphics[width=8.7cm,height=12.4cm]{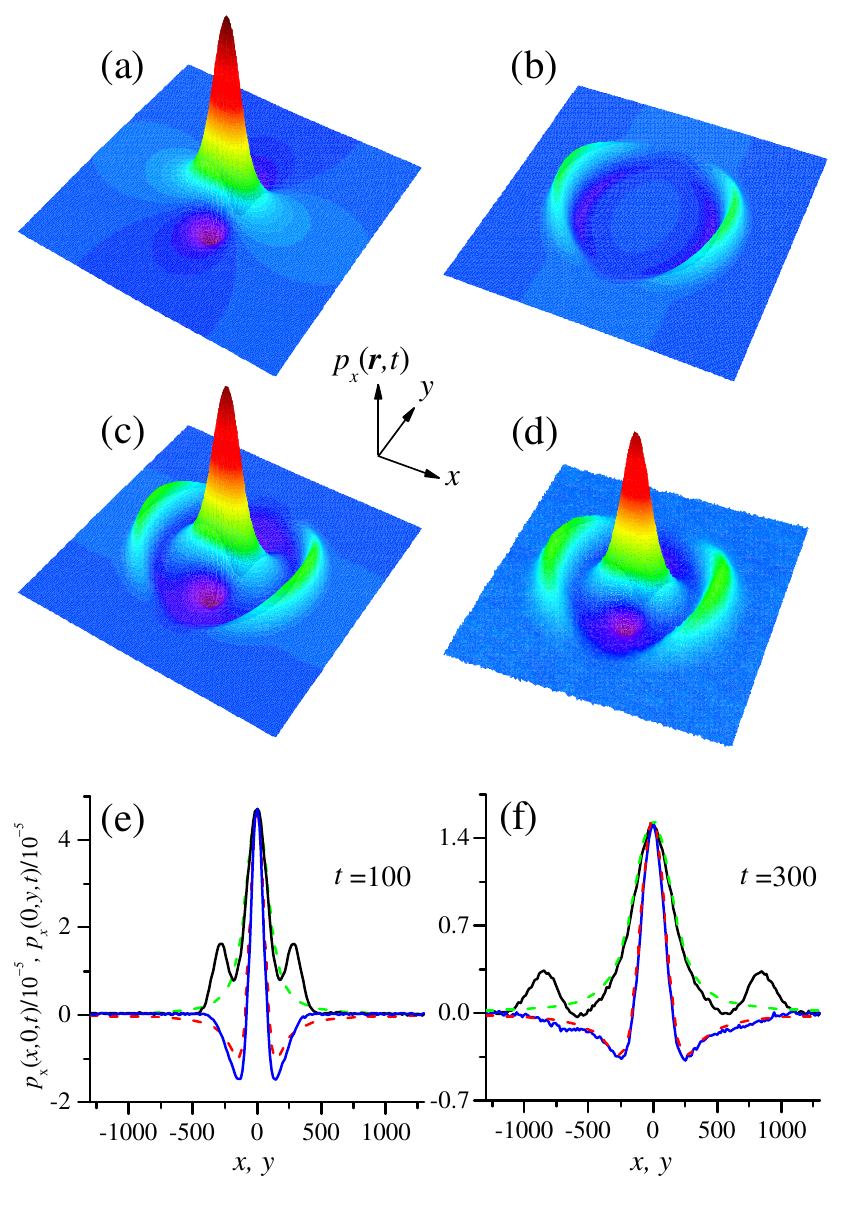}
\caption{The hydrodynamic modes. The analytical results for the viscosity mode (a), the sound mode (b), and their superposition (c) at $t=300$ for $\nu_{0}=8.0$. (d) The simulated $p_{x}(\mathbf{r},t)$ at $t=300$; (e) and (f): The intersections of the simulated $p_{x}(\mathbf{r},t)$ with $y=0$ (black solid lines) and $x=0$ (blue solid lines) at $t=100$ and $300$, respectively. They are compared with the viscosity mode given by Eq. (12) with the best fitting value of $\nu_{0}=8.0$. The intersections of the profiles thereby obtained with $y=0$ ($x=0$) are represented by the green dashed (red dashed) lines. For simulations $\sigma=6$.}
\end{figure}

Note that $p_x(\mathbf{r},t)$ can be obtained following the conventional hydrodynamic approach~\cite{sl}. We take the initial condition $\mathbf{p}(\mathbf{r},t=0)=[p_{x}(0)\delta(\mathbf{r}),0,0]$, $\Delta T(\mathbf{r,}t=0)=\Delta T\delta(\mathbf{r})$, and $\Delta n(\mathbf{r,}t=0)=\Delta n\delta(\mathbf{r})$, where $\Delta T(\mathbf{r,t})$ and $\Delta n(\mathbf{r,t})$ are the temperature and particle density fluctuations, respectively. For this initial condition,  the $5\times5$ hydrodynamic matrix is reduced to a $3\times3$ one, and subsequent analytical derivations are substantially simplified. Solving the hydrodynamic equations (see Supplementary Materials~\cite{a2}, section S1), we obtain $p_{x}(\mathbf{k},t)=p_{x}^{V}(\mathbf{k},t)+p_{x}^{S}(\mathbf{k},t)$ in the Fourier space under the long wave approximation, where
\begin{equation}
\frac{p_{x}^{V}(\mathbf{k},t)}{p_{x}(0)}=\frac{k_{y}^{2}+k_{z}^{2}}{k^{2}}\exp(-\nu_{0}k^{2}t)\label{3}
\end{equation}
is the contribution of the shear viscosity mode and
\begin{equation}
\frac{p_{x}^{S}(\mathbf{k},t)}{p_{x}(0)}=\frac{k_{x}^{2}}{k^{2}}\exp(-\Gamma k^{2}t)\cos(c_{s}kt)\label{4}
\end{equation}
is that of the sound mode. Here $\mathbf{k}$ is the wave number, $\nu_0$ the viscosity diffusivity, $c_{s}$ the sound speed, and $\Gamma$ the sound attenuation coefficient. Note that all these results apply to 2D as well where $k_z=0$.

To get $C_H(t)$ next, we employ the Parseval formula to transform the spatial integral into a wave-vector integral. It gives that
\begin{eqnarray}
\frac{C_H(t)}{C(0)}& = & \frac{M(d-1)}{\rho d}(1-e^{-\frac{C(0)}{D_{0}}t})R(t)\nonumber \\
 &  & [4\pi(D_{H}(t)+D_{0}+\nu_{0})t]^{-\frac{d}{2}},
\end{eqnarray}
where we have neglected the contribution of sound mode as usually adopted~\cite{sl, martin, ernst1}(see Supplementary Materials~\cite{a2}, section S2). Inserting $C_K(t)$ into Eq.~(1), we have $D(t)=D_0[1-\exp(-C(0)t/D_0)]+D_H(t)$, with
\begin{equation}
D_H(t)=\int_{0}^{t}C_H(t^{\prime})dt^{\prime}.
\end{equation}
Solving these two coupled equations, we obtain that
 \begin{eqnarray}
\frac{C_{H}(t)}{C(0)} & = & \frac{M(d-1)}{\rho d}(1-e^{-\frac{C(0)}{D_{0}}t})R(t)\nonumber \\
 &  & (4\pi t)^{-\frac{d}{2}}A^{-\frac{d}{d+2}},\\
D_{H}(t) & = & A^{\frac{2}{d+2}}-(D_{0}+\nu_{0}),
\end{eqnarray}
 with
\begin{eqnarray}
 A&=&(D_{0}+\nu_{0})^{\frac{d+2}{2}}+\frac{M(d-1)(d+2)}{2\rho d}(4\pi)^{-\frac{d}{2}}C(0)\nonumber\\
 & &\int_{0}^{t}(1-e^{-\frac{C(0)}{D_{0}}t^{\prime}})R(t^\prime)t^{\prime-\frac{d}{2}}dt^{\prime}.
\end{eqnarray}

\begin{table}
\begin{tabular}{cccccc}
\hline
$\sigma(\phi)$ & $\nu_{0}$(E)   & $\nu_{0}$(S) & $D_{0}$(E) & $D_{0}$(S)&$\tau$(S)\\
\hline
2(0.03) & 14.1 & $14.3\pm0.05$  & 13.40 & $13.35\pm0.02$ & $13.24\pm0.03$ \\
4(0.13) & 7.7  & $8.3\pm0.05$   & 5.70  & $5.68\pm0.02$  & $5.64\pm0.03$  \\
6(0.28) & 6.3  & $8.0\pm0.05$   & 2.76  & $2.74\pm0.02$  & $2.74\pm0.03$  \\
8(0.50) & 8.5  & $15.0\pm1.0$   & 1.10  & $1.14\pm0.02$  & $1.14\pm0.03$  \\
9(0.63) & 14.3 & $35.0\pm2.0$   & 0.59  & $0.66\pm0.02$  & $0.65\pm0.03$  \\
\hline
\end{tabular}
\caption{Kinetic coefficients obtained by the Enskog formula (E) and by simulations (S).}
\end{table}

To find the first order correction of $R(t)$, we note that the recovery of momentum memory takes place after a ring collision~\cite{cohen1}. The lowest order rings involve three collision events. This process defines the time threshold at which the backflow begins to play a role. The free time follows the Gamma distribution with parameter $\tau$, i.e., $\frac{1}{\tau} \exp(-t/\tau)$. Since $R(t)$ is the probability for the memory recovery to occur, it is the joint probability for three independent collision events to occur. This gives
\begin{equation}
R(t)=1-\frac{\tau_B^2e^{-\frac{t}{\tau_B}}-(2\tau_B\tau_F-\tau_F^2+(\tau_B-\tau_F)t)e^{-\frac{t}{\tau_F}}}{(\tau_F-\tau_B)^2}
\end{equation}
where $\tau_B$ is the mean inverse frequency of the Brownian particle-fluid particle collisions, and $\tau_F$ is that of collisions between fluid particles.

Our result Eq.~(8) is consistent with the standard kinetic theory at $t\rightarrow0$, i.e., $C(t)\approx C_K(t)$ at short times. In contrast, the previous result by the generalized Langevin equation fails in this limit~\cite{martin, zwanzig2, russel}: it leads to  $C(t\rightarrow0)=\frac{k_{B}T}{M_{\ast}}$ with  $M_{\ast}=M+\frac{1}{12}\pi \sigma^{3}\rho$, which should be $C(t\rightarrow0)=\frac{k_{B}T}{M}$ according to the equipartition theorem. Here $k_B$ is the Boltzmann constant and $T$ the temperature. Only for $\rho/\rho_B\ll1$( $\rho_B$ is the density of the Brownian particle), $M_{\ast}\sim M$, it approaches the kinetic result; but for $\rho/\rho_B\sim 1$, a significant deviation arises: $C(t\rightarrow0)=\frac{2}{3}\frac{k_{B}T}{M}$, which remarkably
violates the equipartition theorem. As a consequence, $C(t)$ is smaller than that of the kinetic prediction for $t\approx 0$, leading to a significant underestimation of the diffusion coefficient(See Supplementary Materials~\cite{a2}, section S6).

At long-time limit, for $d=2$, Eq.~(8) and (9) give the asymptotic solution $C_{H}(t)=C(0)\sqrt{\frac{M}{16\pi\rho}}(t\sqrt{\ln t})^{-1}$ and $D_{H}(t)=C(0)\sqrt{\frac{M\ln t}{4\pi\rho}}$ applying for $t\gg t_{c}\equiv\exp[4\pi\rho M(\frac{D_{0}+\nu_{0}}{k_{B}T})^{2}]$, when ($D_{0}+\nu_{0}$) is much smaller than $D_{H}(t)$. For sufficiently large $t$ but smaller than $t_{c}$, the $\sim t^{-1}$ behavior shows up. That is, the $\sim t^{-1}$ behavior appears in intermediate time, and crosses over to $\sim(t\sqrt{\ln t})^{-1}$ behavior at $t_{c}$. Therefore, Eq.~(8) unifies the traditional hydrodynamic and self-consistent solutions~\cite{wain,tlnt,tlnt2,njp}. For $d=3$, our result is $C(t)/C(0)\sim\frac{2}{3}\frac{M}{\rho}[4\pi(D_{0}+\nu_{0}+D_{H}(\infty))t]^{-\frac{3}{2}}$ at long-time limit. Here $D_{H}(\infty)$ is the saturation value of $D_{H}(t)$. This asymptotic behavior differs from previous predictions. Indeed, for the fluid particle, hydrodynamic approaches~\cite{alder1, cohen1} have predicted that $\frac{C(t)}{C(0)}\sim\frac{2}{3}\frac{M}{\rho}[4\pi(D_{0}+\nu_{0})t]^{-\frac{3}{2}}$, while for the Brownian particle, the generalized Langevin equation approach has suggested that $\frac{C(t)}{C(0)}\sim\frac{2}{3}\frac{M_{\ast}}{\rho}(4\pi\nu_{0}t)^{-\frac{3}{2}}$~\cite{martin}. The former corresponds to the neglecting of $D_{H}(\infty)$ in our results, and the latter deviates significantly from ours where there is no room for $D_{0}$ and $M$ being replaced by the effective mass $M_\ast$. Therefore, the previous results at long-time limit are approximations of ours(See Supplementary Materials~\cite{a2}, section S6).

Numerical simulation of a 3D fluid for our aim here is still challenging with available computing resources.    We therefore turn to the 2D hard-disk fluid model to test our results. It consists of $N$ disks of unit mass moving in a square box of size $L$ with the periodic boundary conditions. Considering that the simulation results are free from finite-size effect for $t<L/(2c_{s})$~\cite{pre08}, we set $L=2000$ (see Supplementary Materials~\cite{a2}, section S3). The particle number $N=40000$, corresponding to an average disk number density $n=0.01$. The packing density $\phi=n\pi\sigma^{2}/4$, with $\sigma$ being the disk diameter. The system's behavior at $2 \leq \sigma \leq 9$, which covers from the gas to the liquid regime, are studied in great details. Note that for this model the crystallization density is $\phi=0.71$, corresponding to $\sigma=9.5$. We simulate the system by the event-driven algorithm. The rescaled temperature $T=1$ ($k_{B}$ is set to be unity).

The VACF is only governed by the kinetic parameters $\nu_0$, $D_0$, and $\tau$, which can be obtained analytically for fluid particles by the Enskog formula~\cite{pre06,sl}. However, though a lot of efforts have been devoted to numerically testing the Enskog formula, its accuracy is still to be verified. In the following we will compute these parameters by both the Enskog formula and direct simulations.

To measure $\nu_0$ numerically, we note that the viscosity mode, the inverse Fourier transform of Eq. (3), is
\begin{equation}
\frac{p_{x}^{V}(\mathbf{r},t)}{p_{x}(0)}=\frac{x^{2}-y^{2}}{2\pi r^{4}}(1-e^{-\frac{r^{2}}{4\nu_{0}t}})+\frac{y^{2}}{4\pi r^{2}\nu_{0}t}e^{-\frac{r^{2}}{4\nu_{0}t}}.\label{11}
\end{equation}
The sound mode can be obtained by numerically performing the inverse transform of Eq.~(4). Combining them together leads to a theoretical prediction for $p_{x}(\mathbf{r},t)$. As shown in Fig.~1(a)-(c), $p_{x}(\mathbf{r},t)$ is anisotropic. Note that Eq.~(12) gives the $x$-component of velocity field of vortex backflow. 

\begin{figure*}[!]
\begin{center}
\centerline{\includegraphics[width=17.8cm,height=7.82cm]{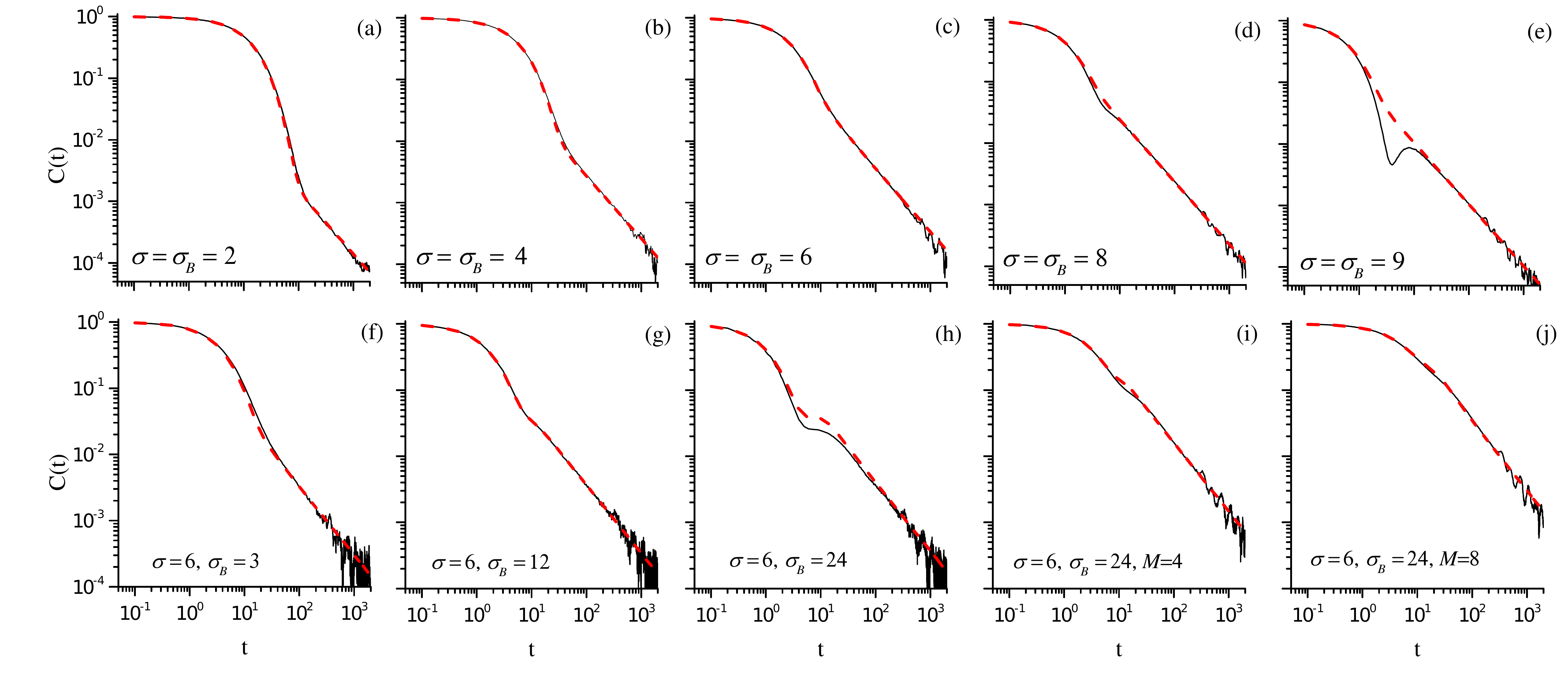}}
\caption{Comparison of the analytical (red dashed line) and the numerical (black solid line) result of the VACF.
For (a)-(h), the mass of the Brownian particles is set to be unity, the same as the fluid particles. In (a)-(e), the diameter of the Brownian particles $\sigma_B$ is the same as the fluid particles as well, and in (f)-(h), $\sigma_B$=3, 12, and  24, respectively. (i)-(j) are for a Brownian particles with diameter $\sigma_B$= 24 and mass $M=4$ and 8, respectively. The parameters $\nu_0$ and $D_0$ for fluid particles are given in Table 1. $D_0$ for Brownian particles in (f)-(j) is obtained by fitting the simulated VACF around $t=0$ (see text) which gives $D_0$=3.85, 1.70, 1.1, 0.85, and 0.80, respectivly.}
\end{center}
\end{figure*}

Numerically, $p_{x}(\mathbf{r},t)$ is computed by the correlation function $\langle \widetilde{p}_{x} (\mathbf{r},t) p_{x}(0) \rangle$, i.e.,
\begin{equation}
\frac{p_{x}(\mathbf{r},t)}{p_{x}(0)}=\frac{\langle\widetilde{p}_{x}(\mathbf{r},t)p_{x}(0)\rangle-\langle\widetilde{p}_{x}(\mathbf{r}\neq0,0)p_{x}(0)\rangle}{\langle|p_{x}(0)|^{2}\rangle}.\label{12}
\end{equation}
Here $\widetilde{p}_{x}(\mathbf{r},t)$ represents the $x$-component of the instantaneous momentum density. The result is shown in Fig. 1(d), which confirms the theoretical prediction.

The explicit expression of the viscosity mode allows us to compute $\nu_{0}$ based on the simulated $p_{x}(\mathbf{r},t)$. Specifically, we fit the viscosity mode, i.e., the center peak of the simulated $p_{x}(\mathbf{r},t)$, by $p_{x}^{V}(\mathbf{r},t)$. In this way we find $\nu_{0}=8.0$ for $\sigma=6$. Note that this value is independent of time [c.f. Fig. 1(e)-(f)], implying that hydrodynamics is harmless to the viscosity diffusivity. Previous numerical studies based on the Einstein-Helfand formula have shown that the viscosity diffusivity is independent of the system size either~\cite{pre06}, consistent with our findings here. In Table 1 we summarize the value of $\nu_{0}$ computed for various packing densities (the state equation used in the Enskog formula is the Henderson expression). It is important to note that the result is close to that given by the Enskog formula under the first Sonine polynomial approximation~\cite{pre06} in the dilute gas regime, but the discrepancy between them becomes significant as the packing density increases. 

We measure $D_0$ by fitting the simulated VACF with $C_K(t)$ at $t\rightarrow0$. We emphasize that the time window for fitting must be sufficiently narrow to ensure that the hydrodynamic effect does not play a role yet (see Supplementary Materials~\cite{a2}, section S4, S5). The results for a fluid particle are given in Table 1. As a comparison, the $D_0$ values obtained by the Enskog formula are given in Table 1 as well: It can be seen that the agreement between them is excellent for almost all densities: the discrepancy is less than $10\%$ even for the highest density ($\sigma=9$) studied in simulations. For the diffusion constant $D_0$, contrary to the common belief that the Enskog formula is only accurate at low densities, our results suggest that it is valid at a much broader regime. In fact, by employing the relation between the mean free time $\tau$ and the kinetic diffusion constant~\cite{sl}, i.e., $D_0 = \frac{C(0)d}{2}\tau $,  we can calculate $D_0$ in the third way. In a 2D fluid, it appears $D_0=\tau$ with dimensionless unity. The parameter $\tau$ can be easily measured in simulations, and the result is given in Table 1. The mean free time is determined only by the first collision between tagged particle and fluid particles, in which case there is no memory feedback taking place. It thus describes the pure kinetic effect.   The results are in perfect agreement with that by fitting the short-time behavior of the VACF, confirming again the validity of the latter.

With $D_0$, $\nu_0$ and $\tau$ at hand, we can calculate $C(t)$ and $D(t)$. Figure 2(a)-(e) show the VACF for a fluid particle at different packing densities, Fig. 2(f)-(h) are for a Brownian particle of unit mass with different diameters, and Fig. 2(i)-(j) are for a Brownian particle of the fixed diameter but different masses. We see that in most cases Eq.~(7) is in excellent agreement with the simulation results. A slight discrepancy between theoretical and numerical results appears in certain extreme conditions, as in Fig.~2(e) and (h). For example, In Fig.~2(e) the fluid density is so high that it is very close to the crystallization point that the system may have deviated from the fluid structure to certain extent. Fig. 2(h) is for a Brownian particle with $\rho/ \rho_B \sim5.8$, whose density is much smaller than the fluid. These extreme situations have been far beyond the usual scope of Brownian motion. In spite of this fact, it is encouraging that the deviation in the diffusion coefficient is negligible (see Supplementary Materials~\cite{a2}, section S6). 

In summary, we have derived a general formula for Brownian motion, applicable to a broad spectrum of Brownian particles regardless of their shapes and sizes (if the rotation is ignored); and to very general environments, ranging from dilute gases to dense liquids. Our result also allows us to describe the kinetic-hydrodynamic crossover of the VACF, leading to an accurate account of both the kinetic and hydrodynamic contributions to the diffusion coefficient. For 3D case, the formula of generalized Langevin equation approach is   a crude approximation to ours and fails to quantitatively calculate the diffusion coefficient for light particles, whereas our formula indicates that the hydrodynamic contribution can reach the order of the kinetic diffusion constant. For 2D cases, our formula gives, for the first time, the full expression of the diffusion coefficient, which indicates that the hydrodynamic effect may dominate. Finally, our study also suggests an accurate way to measure the kinetic diffusion constant and the viscosity diffusivity. By comparison, we find that the Enskog formula is very accurate for estimating the diffusion constant, but applies only at low density for the viscosity diffusivity.  

\begin{acknowledgments}
We are grateful to J. Wang and C. Tian for useful discussions. This work is supported by the NSFC (Grants No. 11335006).
\end{acknowledgments}

\bibliography{bibfile}

\end{document}